\documentclass[sigconf]{acmart}
\usepackage{graphicx} 
\usepackage{xcolor}
\usepackage{tcolorbox}
\usepackage{fontawesome}
\newcounter{FindingCounter}
\NewTColorBox{custombox}{o}{
    colback=cyan!8,
	colframe=cyan!60,
	left=1.0mm,
	right=1.0mm,
	top=0.3mm,
	bottom=0.3mm,
	fonttitle=\bfseries,
	arc=0mm,
	leftrule=0mm,
	rightrule=0mm,
	toprule=0mm,
	bottomrule=0mm,
	notitle,
	before=\par\medskip\noindent,
    IfValueTF={}{before upper={\faHandORight{}~\textbf{New Idea~\#\arabic{FindingCounter}\stepcounter{FindingCounter}: } }}{},
    parbox=false,
}

\begin{document}
\title{Trustworthy AI Software Engineers}
\author{Aldeida Aleti}
\affiliation{%
  \institution{Monash University}
  \country{Australia}
}
\email{aldeida.aleti@monash.edu}

\author{Baishakhi Ray}
\affiliation{%
  \institution{Columbia University}
  \country{USA}
}
\email{rayb@cs.columbia.edu}

\author{Rashina Hoda}
\affiliation{%
  \institution{Monash University}
  \country{Australia}
}
\email{rashina.hoda@monash.edu}

\author{Simin Chen}
\affiliation{%
  \institution{Columbia University}
  \country{USA}
}
\email{sc5687@columbia.edu}

\begin{abstract}
With the rapid rise of AI coding agents, the fundamental premise of what it means to be a software engineer is in question. In this vision paper, we examine what it means for an AI agent to be considered a software engineer and then critically think about what makes such an agent trustworthy. Grounded in established definitions of SE (SE) and informed by recent research on agentic AI systems, we conceptualise AI software engineers as participants in human-AI SE teams composed of human software engineers and AI agents, and we distinguish trustworthiness as a key property of these systems and actors rather than a subjective human attitude. Extending on historical perspectives and emerging visions, we identify key dimensions that contribute to the trustworthiness of AI software engineers, spanning technical quality, transparency and accountability, epistemic humility, and societal and ethical alignment. Beyond defining these dimensions, we address a critical but underexplored challenge: how trustworthiness can be operationalised in practice. We therefore introduce the notion of evidence-centric inspection, arguing that developers should evaluate selective signals and justifications of trustworthiness rather than raw outputs, and we outline implications for rethinking verification, validation, and code review in human-AI SE teams. 
\end{abstract}
\maketitle

\section{Introduction}

Software engineering (SE) is undergoing a disruptive transformation driven by the rapid emergence of Large Language Model (LLM)-based agents capable of generating, repairing, and analysing code. Recent advances in agentic systems such as Claude Code~\cite{anthropic_claude_code_overview}, Codex~\cite{openai_codex_github} and Kiro~\cite{kirodotdev_kiro_github} and some of the earlier generation agents including AutoCodeRover~\cite{zhang2024autocoderover}, SWE-agent~\cite{yang2024swe}, RepairAgent~\cite{bouzenia2025repairagent} and RovoDev~\cite{tantithamthavorn2026rovodev}
demonstrate the feasibility of increasingly autonomous Artificial Intelligence (AI) coding agents.
These systems promise productivity gains by automating tasks that traditionally require substantial human effort~\cite{ziegler2024measuring}. Yet, despite significant technical progress, their adoption in practice is tempered by persistent skepticism regarding their trustworthiness~\cite{Roychoudhury26}. Empirical evidence substantiates these concerns. Prior work shows that LLMs can be untrustworthy~\cite{turpin2023language,gong2024well}, misunderstand vulnerabilities~\cite{ullah2024llms}, hallucinate~\cite{wang2024contracttinker}, memorise training data~\cite{kong2025demystifying,dycodeeval}, rely on biased features~\cite{turpin2023language}, or generate syntactically correct yet semantically invalid programs~\cite{wang2025towards,tan2024prompt}. These behaviours impact trust in coding agents.

Addressing trustworthiness of AI software engineers first requires clarifying what it means to be a software engineer in the age of AI. Established definitions from IEEE, ACM, and the SE Body of Knowledge (SWEBoK) characterise SE as far more than programming \cite{hoda2026toward}. Empirical studies of professional practice consistently show that a substantial portion of day-to-day work involves non-coding activities such as eliciting and negotiating requirements, analysing specifications, reviewing and maintaining code, coordinating with stakeholders, and documenting design decisions~\cite{Meyer21}. It is unclear how present day and emerging AI coding agents can account for socio-technical activities of SE beyond coding. These activities require judgment, communication, and accountability, among other social constructs. As AI capabilities advance, the role of human engineers is shifting toward supervision, orchestration, validation, and ethical oversight, raising fundamental questions about the future of human–AI SE teams. To address this, we ask a fundamental question: \textit{\textbf{What is an agentic engineer?}} 

Agentic software engineers offer the promise of automation, scalability, and productivity gains. At the same time, the risks of uncritical automation, i.e., automation without adequate understanding of its risks and limitations, are substantial. If agentic software engineers operate without clear behavioural boundaries, accountability mechanisms, or ethical constraints, failures may become difficult to diagnose and responsibility and accountability may be obscured. These tensions highlight the need for agentic engineers that are not only capable, but also trustworthy. This leads to a second fundamental question: \textit{\textbf{What characteristics should agentic engineers possess in order to be considered trustworthy?}} 

Trustworthiness extends beyond raw performance as measured by lines of code, and previous work has introduced dimensions such as fairness, robustness, transparency, generalization, explainability, auditability, traceability, alignment with human values, and privacy~\cite{li2023trustworthy}. Although these dimensions are widely discussed in the broader trustworthy AI literature, they have not yet been systematically grounded in the context of AI-based SE agents. Khati et al. \cite{khati2025mapping} conceptualized trust in the context of SE through a systematic literature review and identified three key attributes of trustworthiness for coding models: \textit{ability} (measured in terms of accuracy, efficiency, and reliability), \textit{benevolence} (capturing alignment with user and stakeholder goals), and \textit{integrity} (relating to safe, secure, and ethical development practices). While this provides an important foundation for understanding trust in AI-based coding support, agentic systems extend beyond conventional coding models in both scope and responsibility. Unlike tools that primarily generate or complete code, agentic systems are expected to \textit{plan} tasks, \textit{interact} with tools and environments, \textit{make} intermediate decisions, and \textit{collaborate} with human teammates across the SE lifecycle. As a result, trustworthiness in agentic SE systems must be understood more broadly: not only in terms of the quality of their outputs, but also in terms of the soundness of their decision-making processes, their transparency and accountability, their ability to operate safely under uncertainty, and their capacity to remain aligned with human intentions, team norms, and organizational constraints.

Identifying dimensions of trustworthiness alone is insufficient to support their practical adoption. As agentic systems generate increasing volumes of code, it becomes infeasible for human developers to inspect everything. Requiring exhaustive human validation risks creating cognitive overload and review fatigue, ultimately limiting the effectiveness of human-AI SE teams.  Developer fatigue from reviewing AI generated work is also a rising concern. This leads to a third fundamental question: \textit{\textbf{What should developers verify and validate when working with agentic software engineers?}} Addressing this question requires moving beyond abstract notions of trustworthiness toward actionable practices that enable developers to selectively inspect, interpret, and rely on AI-generated artefacts. In this paper, we argue that this challenge necessitates a shift from artefact-centric inspection toward evidence-centric inspection, where developers evaluate signals and justifications of trustworthiness rather than raw outputs.




This vision paper aims to move the discourse beyond productivity-centric narratives of AI in SE, much of which is explicitly or implicitly focused on AI replacing human software engineers, towards a principled understanding of trustworthy human-AI SE teams.

\section{What is an Agentic Engineer?}
Software Engineering (SE) extends beyond coding \cite{hoda2026toward}, hence an agentic engineer must go beyond coding and coding-related activities. Although current AI agents demonstrate impressive proficiency in code generation, repair, and refactoring, SE encompasses a broader range of activities, including requirements gathering (e.g., writing user stories), requirements analysis (e.g., analysing specifications for ambiguity or inconsistency), architectural design, testing, and verification, and long-term maintenance and evolution. Moreover, SE is a collaborative endeavour, \textit{a team effort}. Recently, there has been a shift in the SE community from viewing AI agents as isolated coding tools toward treating them as members of a \emph{SE team} \cite{abrahao2025software, hassan2025agentic, roychoudhury2025agentic, hoda2026toward}. A set of \textit{comprehensive}, \textit{responsible}, \textit{adaptive}, \textit{foundational}, \textit{translational} (CRAFT) values and principles have been proposed for agentic SE that expand its focus beyond coding activities and emphasize human-AI collaboration \cite{hoda2026toward}. An agentic engineer must therefore be able to work not only independently, but also in coordination with other AI agents and with human software engineers, supporting a human-AI collaboration model rather than human replacement, akin to emerging paradigms such as the AI co-scientist~\cite{moons2025google}.
An AI agent should be considered an AI software engineer only if it can assume responsibilities consistent with real-world SE practice: working in teams on socio-technical tasks spanning the SE lifecycle, under constraints and accountability~\cite{donta2025socio, dam2025towards}. 
Recent SE visions emphasise that while AI is reshaping the field, humans remain essential for quality, governance, and value-sensitive decisions (e.g., Abrahão et al.~\cite{abrahao2025software}). 
Complementary vision work argues that agentic SE cannot be defined solely in terms of code, but must reflect SE foundations, values, and processes~\cite{hoda2026toward}. 
Grounded in these perspectives and recent evidence on AI’s strengths and limits in realistic SE workflows~\cite{murali2024ai}, we identify four defining characteristics of an agentic software engineer.

\begin{custombox}[ ]
For an AI agent to be considered an agentic engineer, it must satisfy four conditions.

\textbf{i)} \textit{The agent must be capable of handling a range of SE tasks beyond coding} such as evolving problem statements, requirements, constraints, ethics, design, testing and operations, requiring iterative refinement under partial information~\cite{fakhoury2024llm}.

\textbf{ii)} \textit{The agent must demonstrate agency through planning and tool use}, extending beyond code-centric tasks~\cite{hoda2026toward,sapkota2025ai}. 

\textbf{iii)} \textit{The agent must be collaborative}. As SE is fundamentally team-based, an agentic engineer must integrate into human workflows by supporting coordination, feedback, and negotiation~\cite{murali2024ai,abrahao2025software,hoda2026toward}.

\textbf{iv)} \textit{The agent must respect human values, constraints, and ethical responsibility}, It must justify decisions, and support accountability and oversight, consistent with ethics-by-design and sustainability principles~\cite{murali2024ai,hoda2026toward}.
\end{custombox}

This working definition intentionally sets a higher bar than ``autonomous coding'' or AI coding agents. It matches emerging evidence that real-world usefulness depends on interaction, context, validation, and integration into socio-technical workflows~\cite{murali2024ai}, and it aligns with the push to frame agentic SE as beyond code, grounded in SE foundations, values, and shared vocabulary~\cite{hoda2026toward}

\section{What Makes an Agentic Engineer Trustworthy?}
Trust is a central construct in human-AI collaboration, yet it is conceptually distinct from trustworthiness \cite{pink2025trust}. Trust refers to a human decision or disposition to rely on an AI system in a given context, while trustworthiness denotes the properties of the system that justify such reliance \cite{pink2025trust}. The two are related but asymmetrical: humans may place trust in untrustworthy systems, and conversely, trustworthy systems may fail to be trusted if their behaviour or intentions are opaque. This distinction is particularly important in SE, where inappropriate trust can lead to severe technical, organisational, or societal harm, while insufficient trust may prevent realising productivity benefits. We focus on trustworthiness as a system property rather than on subjective trust alone.

Prior studies of trust describe trust as a dynamic, living concept, one that can be gained, lost, or recalibrated, sometimes abruptly in response to failures or incidents~\cite{salehi2024trust,yang2023toward,pink2025trust}. Roychoudhury et al.~\cite{roychoudhury2025agentic} argue that the key barrier to the adoption of AI software engineers is not organisational resistance but developer trust, shaped by day-to-day interactions with AI agents in real workflows. Empirical work further suggests that developers approach AI systems with different default trust positions, shaped by experience, role, and values, rather than a uniform baseline of skepticism or optimism~\cite{spiegler2025images}. These observations suggest that trustworthiness cannot be established once and for all; instead, it must support sustained, recalibrated trust relationships over time, including notions of mutual trust between human developers and AI agents.

Importantly, no single dimension is both necessary and sufficient for trustworthiness. It is unlikely that any one dimension alone can guarantee trustworthiness, nor that the absence of a single dimension always makes a system untrustworthy in all contexts. Rather, trustworthiness emerges from configurations of dimensions that collectively justify reliance in a given setting. This motivates a multidimensional view that avoids oversimplification while still providing structure. 

\begin{figure} [!ht]
\vspace{-3mm}
    \centering
    \includegraphics[width=\linewidth]{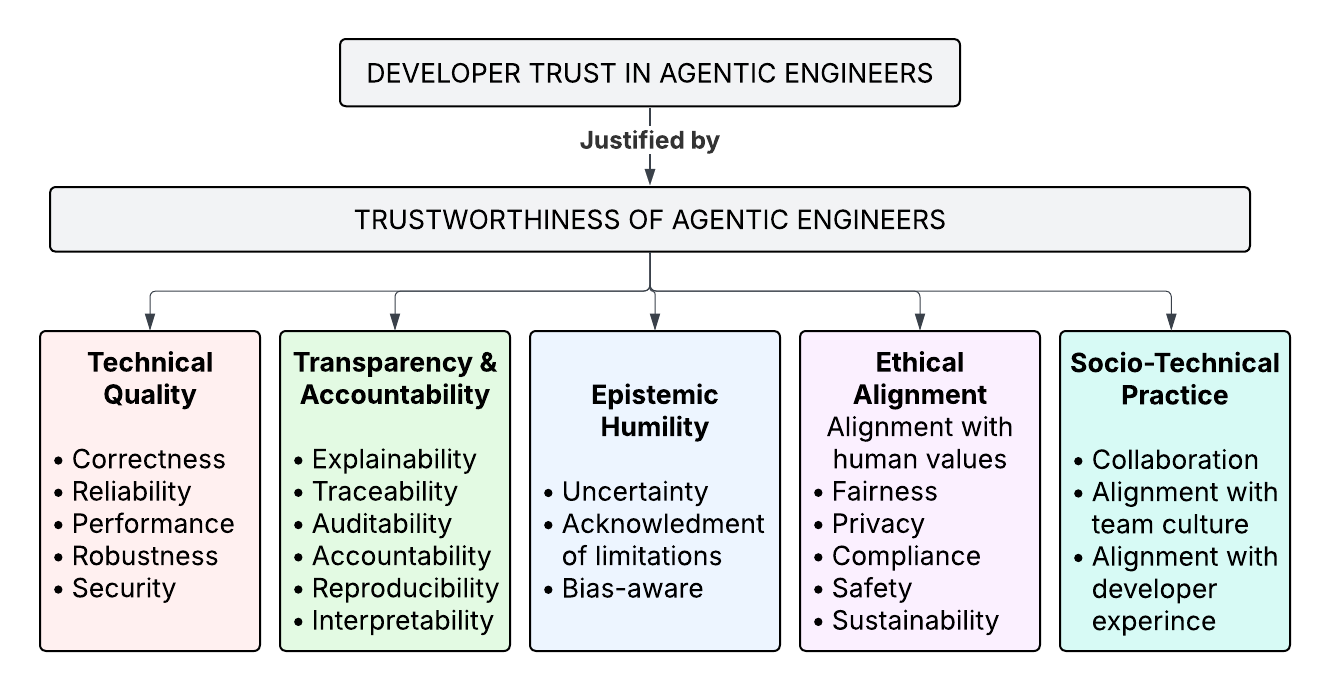}
    \caption{Multidimensional View of Trustworthiness in Agentic Engineers.}
    \label{fig:dimensions}
    \vspace{-2mm}
\end{figure}

\begin{custombox} \textbf{A multi-dimensional trustworthiness framework for agentic engineers.} we identify several key dimensions that are directly related to the trustworthiness of AI software engineers, as shown in Figure~\ref{fig:dimensions}. These dimensions apply at different levels (system, model, and output) and matter to different stakeholders (developers, organisations, and society), often over both short- and long-term horizons.
They are grounded and extend prior work in trustworthy AI~\cite{Roychoudhury26} and with grounding on SE quality models (e.g., SWEBoK), and recent work on trustworthiness in AI-assisted development~\cite{tung2024automated}. 
\end{custombox}

\textit{Technical quality is foundational to trust.} An AI software engineer cannot be considered trustworthy if it frequently produces incorrect results or behaves inconsistently. Correctness refers to whether outputs meet expected functional requirements, while reliability concerns the system’s ability to deliver acceptable results consistently over time. These dimensions are often assessed using quantitative metrics at the model or output level, but their implications extend to system-level trust in engineering workflows.
Performance and robustness further shape technical quality. Performance includes response time, efficiency, scalability, and resilience to denial-of-service-like behaviours, while robustness concerns correct behaviour under invalid inputs or stressful conditions. Security is also critical, ensuring that systems resist misuse and vulnerabilities. In addition, cost—spanning developer time, computational resources, financial expense, and environmental impact-affects whether systems remain practical and trustworthy at scale. Even a correct system may be untrustworthy if it is prohibitively slow, fragile, or resource-intensive.

\textit{Transparency and accountability enable responsible reliance.} Transparency and explainability concern whether stakeholders can understand how and why decisions are made. Traceability enables linking outputs to inputs, models, and processes, while auditability and reproducibility support verification. Accountability requires that actions and decisions can be justified to different audiences, ensuring that responsibility is clearly assigned and enforceable, while reproducibility refers to obtaining consistent results under the same conditions.
\textit{Epistemic humility captures the limits of AI knowledge.} AI software engineers must communicate uncertainty, acknowledge limitations, and remain aware of potential biases. Without such humility, systems risk overconfidence, misuse, and misplaced trust, particularly in complex or high-stakes settings.
\textit{Ethical alignment ensures consistency with human values~\cite{rokeach2008understanding}.} This includes fairness, bias mitigation, privacy protection, regulatory compliance, safety, and sustainability. These dimensions are critical given the societal reach of software systems. Failures here may not be immediately visible at the code level, yet they can significantly erode trust over time.
\textit{Trustworthiness is also shaped by ethical alignment and socio-technical practice.} Beyond individual systems, trust depends on how AI software engineers integrate into human teams and broader contexts. Collaboration, alignment with team culture, and support for developer experience influence adoption and effective use. More broadly, societal impact, cultural sensitivity, and sustainability shape whether these systems are perceived as responsible participants in socio-technical ecosystems.

\section{What should developers verify and validate when working with agentic engineers?}

Given the key dimensions of trustworthiness introduced in the previous section, a critical open question remains unresolved: what should human developers actually verify and validate when working with agentic software engineers? The trustworthiness dimensions are not always translated into actionable inspection practices in real-world SE workflows. Agentic systems generate increasingly large volumes of code and development artefacts and this gap becomes more pronounced. Human developers cannot feasibly inspect everything, and requiring them to do so risks cognitive overload, review fatigue, and ultimately a breakdown in effective human-AI collaboration. The challenge, therefore, is not only to define trustworthiness, but to make it inspectable under realistic human constraints.

We conceptualise this challenge as a selective visibility problem. Agentic  engineers operate through multi-step processes involving planning, tool use, intermediate decisions, and iterative refinement. Exposing all intermediate artefacts and reasoning steps is neither scalable nor useful for developers. Instead, trust must be supported through carefully curated visibility into the agent's behaviour.

\begin{custombox}[] \textbf{From Artefact-Centric to Evidence-Centric Inspection.}
We propose a shift from artefact-centric inspection, where developers focus primarily on reviewing final outputs such as code, toward evidence-centric inspection, where the goal is to assess whether sufficient justification exists to rely on those outputs. In this view, developers do not need to see everything; rather, they need to see the right things. These include signals about whether the agent’s outputs align with requirements and constraints, whether key decisions are grounded in defensible reasoning, and whether uncertainty, assumptions, and potential failure modes are explicitly communicated. Trustworthiness, therefore, becomes a property not only of what the system produces, but of what it chooses to reveal.
\end{custombox}

This shift in perspective has implications for how verification and validation (V\&V) are understood in the context of agentic engineers. Traditionally, verification concerns whether a system has been built correctly, and validation concerns whether the right system has been built. In human-AI settings, these notions must be extended to account for the processes through which artefacts are generated. Developers must reason not only about the correctness of outputs, but also about the soundness of the agent’s decision-making processes and its interpretation of human intent. This introduces additional layers of verification, including process-level verification, which concerns whether the agent’s planning and tool use are appropriate, and epistemic verification, which concerns whether the agent accurately represents its own uncertainty and limitations. Validation similarly extends beyond outputs to include whether the agent’s behaviour remains aligned with evolving requirements and constraints. As a result, V\&V becomes a multi-level activity spanning outputs, processes, and context. Supporting this form of verification requires new theory, techniques and tools that are explicitly focused for developer-facing inspection of trustworthiness. While traditional approaches such as testing and static analysis remain necessary, they are insufficient in isolation because they operate primarily at the level of final artefacts. In contrast, agentic systems require mechanisms that can summarise, structure, and expose relevant aspects of their internal processes and decisions.

\begin{custombox}[] \textbf{Rethinking Verification and Validation.} 
One promising direction is the development of techniques that distil large volumes of agent activity into concise, meaningful representations that highlight key decisions, assumptions, and risks. Traceability mechanisms can further support this by linking generated artefacts back to requirements, inputs, and intermediate steps, enabling developers to reconstruct how a particular outcome was produced. Additionally, systems that explicitly surface uncertainty, alternative solutions, or known limitations can help developers calibrate their trust more effectively. Emerging work on automated trustworthiness oracles~\cite{tung2024automated} suggests that it may be possible to automatically generate checks for properties such as robustness, fairness, or security, thereby guiding developers toward the aspects of a system that warrant closer inspection.
\end{custombox}

Taken together, these directions point toward a new class of tooling that does not simply analyse code, but actively mediates the interaction between human developers and agentic systems by shaping what is visible, explainable, and verifiable.

These challenges are particularly acute in the context of code review. Code review is inherently a socio-technical activity that involves not only identifying defects, but also understanding design intent, sharing knowledge, and coordinating team practices. With the introduction of agentic systems, the nature of code review is fundamentally altered. The volume of generated artefacts increases substantially, making exhaustive manual inspection impractical, while the opacity of AI-generated outputs makes reasoning about their correctness more difficult. As a result, code review risks becoming a bottleneck in human-AI teams. 

\begin{custombox}[] \textbf{Code Review as Runtime Monitoring.} We propose that in human-AI software engineering teams, code review should extend beyond pre-merge inspection and become a continuous, runtime activity. Deployed code is not considered final, but provisional. AI agents continuously monitor system behaviour in production, evaluating execution traces, resource usage, security signals, and deviations from expected behaviour. This reframes code review as a continuous assurance mechanism embedded within the system lifecycle, where trustworthiness is established through sustained behavioural validation. As a result, review becomes temporally distributed, adaptive, and tightly coupled with system operation, fundamentally dissolving the boundary between development and deployment.

\end{custombox}

This perspective highlights a key research gap: the need to translate abstract dimensions of trustworthiness into concrete, developer-facing interfaces and workflows. While prior work has identified what properties trustworthy systems should exhibit, significantly less attention has been paid to how these properties are operationalised in practice, particularly in the context of human–AI collaboration.

\begin{custombox}[] \textbf{From Trustworthiness Properties to Trust Interfaces.} Designing effective trust interfaces requires understanding not only what information to present, but also when and how to present it in ways that align with developer workflows and cognitive constraints. Without such mechanisms, human–AI SE teams face a fundamental trade-off between exhaustive verification, which is infeasible at scale, and uncritical reliance, which risks misplaced trust. Bridging this gap is essential for enabling agentic software engineers to be not only technically capable, but meaningfully trustworthy in practice.
\end{custombox}

\section{Conclusion}
This paper defined key requirements for AI agents to qualify as agentic software engineers and introduced key dimensions underpinning their trustworthiness. Beyond defining these dimensions, we addressed a critical and underexplored challenge: how trustworthiness can be operationalised in practice. In particular, we posed the question of what developers should verify and validate when working with agentic software engineers. We argued that the scale and complexity of AI-generated artefacts render exhaustive inspection infeasible, and that trustworthiness must instead be made inspectable through selective, developer-facing evidence. To this end, we introduced the notion of evidence-centric inspection and outlined implications for rethinking verification and validation as multi-level activities spanning outputs, processes, and context, as well as for evolving code review into a computationally supported, human–AI collaborative practice.
We further argued that trustworthiness emerges from configurations of dimensions in context, not from any single property. Accordingly, we advocate for continuous, context-sensitive assessment and identify the trustworthiness measurement gap as a key challenge. We call for future empirical and design-oriented research on trust-aware development practices, verification strategies, and human-AI collaboration models to enable appropriate and scalable trust in agentic SE systems.

\section{Data Availability}
There are no data available yet.

\bibliographystyle{acm}
\bibliography{bibliography,cm}
\end{document}